\documentclass[12pt]{iopart}
\usepackage{graphicx}
\usepackage{iopams}

\usepackage{color}
\newcommand{\zk}[1]{#1} 

\begin{document}

\title[From discrete to continuous percolation in dimensions 3 to 7]{From discrete to continuous percolation\\ in dimensions 3 to 7}
\author{Zbigniew Koza, Jakub Po{\l}a}
\address{Faculty of Physics and Astronomy, University of Wroc{\l}aw, 50-204 Wroc{\l}aw,
Poland}
\ead{zbigniew.koza@uwr.edu.pl}

\date{\today}

\begin{abstract}
We propose a method of studying the continuous percolation of aligned objects as a limit of a corresponding discrete model.
We show that the convergence of a discrete model to its
continuous limit is controlled by a power-law dependency with a universal exponent $\theta = 3/2$.
This  allows us to estimate the continuous percolation thresholds
in a  model of aligned hypercubes in dimensions  $d = 3,\ldots,7$
with  accuracy far better than that attained using any other method before.
We also report improved values of the  correlation length critical exponent $\nu$
in dimensions $d = 4,5$ and the values of several universal wrapping probabilities for $d=4,\ldots,7$.
\end{abstract}




\maketitle

\section{Introduction\label{sec:intro}}

While  advances in two-dimensional (2D) percolation have recently allowed to determine
the site percolation threshold  on the square lattice
with an astonishing accuracy of 14 significant digits \cite{Jacobsen2015}
and many critical exponents in 2D have been known exactly for decades \cite{Stauffer1994},
the progress in higher dimensions is far slower.
The main reason for this is that \zk{the two theoretical concepts that proved particularly fruitful in percolation theory,
conformal field theory and duality, are} useful only in 2D systems, and the \zk{thresholds} in higher dimensions \zk{are known only} from simulations.
The site and bond percolation thresholds in dimensions $d=3,\ldots, 13$ are known
with accuracy of at least 6 significant digits
\cite{Grassberger03,Xu2014,Wang2013}, but for more complicated lattices,
e.g. fcc, bcc or diamond lattices \cite{Marck98},
complex neighborhoods \cite{Malarz15}, or  continuum percolation models \cite{Torquato12b,Jin15,Baker2002}
this accuracy is often far from satisfactory. Moreover,
even though the upper critical dimension is known to be $d_\mathrm{u}=6$ \cite{Christensen2005},
numerical estimates of the critical exponents for $d=4,5$
are still rather poor.

Continuous percolation of aligned objects  can be regarded as
a limit of a corresponding discrete model. Using this fact,
we recently improved the accuracy of numerical estimates of  continuous percolation of aligned cubes ($d=3$)
\cite{Koza14}. We also generalized the excluded volume approximation \cite{Balberg1984,Balberg1987}
 to discrete systems and found that the limit of the continuous percolation is controlled by a power-law dependency
with an exponent $\theta =3/2$ valid for both $d=2$ and $3$.
The main motivation behind the present paper is to verify whether the relation $\theta =3/2$
holds also for higher dimensions and if so, whether it can be used
to improve the accuracy of  continuous percolation thresholds
in the model of aligned hypercubes in dimensions $d = 3,\ldots,7$.
With this selection, the conjecture will be verified numerically for all dimensions  $d \le d_\mathrm{u}$
as well as in one case above $d_\mathrm{u}=6$, which should render its generalization to all $d$ plausible.

Answering these questions required to generate a lot of data, from which several other physically interesting
quantities could also be determined.
In particular, we managed to  improve the accuracy of the  correlation length  critical exponent $\nu$ in dimensions $d=4,5$
and to determine the values of various universal wrapping probabilities in dimensions $d=4,\ldots,7$.


\section{The Model\label{sec:Model}}
We consider a hypercubic lattice of the linear size $L$ lattice units (l.u.)\ in a space dimension $d$.
This lattice is gradually filled with hypercubic  ``obstacles'' of linear size $k$ l.u.\ ($k/L \ll 1$)
until a wrapping percolation has been found (for the sake of simplicity,
henceforth we will assume that $L$, $k$ are dimensionless integers).
The obstacles, aligned to the underlying lattice and with their edges coinciding
with  lattice nodes, are deposited at random into the lattice and the periodic boundary conditions
in all directions are assumed to reduce finite-size effects.
During this process the deposited hypercubes are free to overlap;
however, to enhance the simulation efficiency,
no pair of obstacles is allowed to occupy exactly the same position.

\begin{figure}
  \includegraphics[width=0.9\columnwidth]{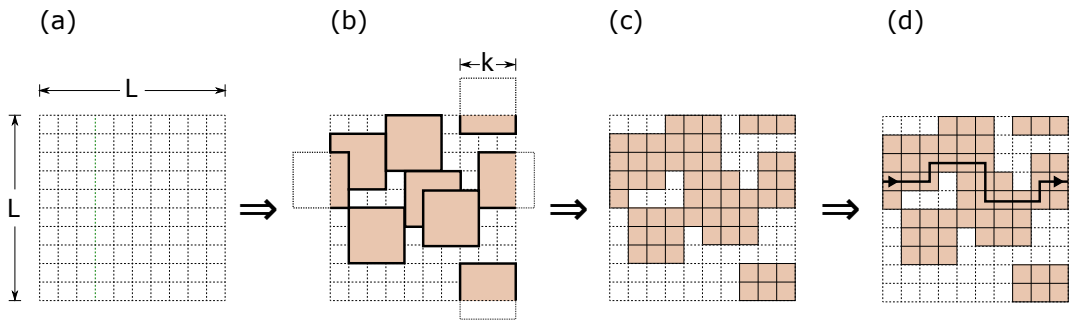}
\caption{\label{fig:model}
 Construction of the model in the space of dimension $d=2$. An empty regular lattice of size $L\times L$
lattice units (l.u.)\ with periodic boundary conditions (a) is filled at random with
square obstacles of size $k\times k$ l.u.\ aligned to the lattice axes (b) and the elementary
cells occupied by the obstacles are identified (c);
finally,  a wrapping path through the occupied elementary cells (site percolation) is looked for (d).
The same method was used for larger $d$.
 }
\end{figure}

As illustrated in figure~\ref{fig:model},
the volume occupied by the obstacles can be regarded as a simple union of elementary lattice cells
and the model is essentially discrete.
Two elementary cells are considered to be connected
directly if and only if they are occupied by an obstacle and share the same hyperface of an  elementary cell.
We define a percolation cluster as a set of the elementary cells
wrapping around the system through a sequence of directly connected elementary cells.
Thus, the model interpolates between the site percolation on a hypercubic lattice  for $k=1$
and the model of continuous percolation of aligned hypercubes
\cite{Torquato12b,Baker2002,Mertens2012} in the limit of $k\to\infty$.

The percolation threshold is often expressed in terms of the volume fraction $\varphi$
defined as the ratio of the number of the elementary cells occupied by the obstacles to the
system volume, $V\equiv L^d$.
What is the expected value of $\varphi$ after $N$ hypercubes have been placed at random (but different) positions?
To answer this question, notice that while the obstacles can overlap, they can be located at exactly
$L^d$ distinct locations and so  $0\le N \le V$.
Moreover, owing to the periodic boundary conditions, any elementary cell can be occupied by exactly $v$
different hypercubes, where $v\equiv k^d$ is the volume of a hypercube. Thus, the probability that
an elementary cell is not occupied by an obstacle, $1-\varphi$, is equal to the product of $v$ probabilities that
no hypercubes were placed at $v$ locations. This implies that
\begin{equation}
  \label{eq:def-phi}
   \varphi = 1 - \left( 1-\frac{N}{V} \right)^v = 1 - \left( 1-\frac{N}{L^d} \right)^{k^d}.
\end{equation}
For $k=1$ this formula reduces to $\varphi = N/V$ irrespective of $d$. In the limit of $k\to\infty$ equation (\ref{eq:def-phi})
reduces to $\varphi = 1 - \exp(-\eta)$, where $\eta =Nv/V = N k^d/L^d$ is the reduced number density  \cite{Baker2002}.


\section{Numerical and mathematical details}\label{sec:numerical-details}
The number of lattice sites in a cluster of linear size $L$ is of order of $L^d$,
a quantity rapidly growing with $L$ in high dimensions $d$.
This imposes severe constraints on numerical methods.
On the one hand, one would like to have a large $L$ to minimize finite-size effects,
which are particularly important near a critical state;
on the other hand, dealing with $L^d$ objects exerts a pressure on the computer storage
and computational time. To mitigate this problem, special algorithms were
developed that focus on the efficient use of the computer memory.
For example, Leath's algorithm \cite{Leath1976},
in which a single cluster is grown from a single-site ``seed'',
turned out very successful in high-dimensional simulations of site and bond percolation \cite{Grassberger03,Stauffer2000}.
However, the use of such algorithms in the present model would be impractical,
as the obstacle linear size $k$ is now allowed to assume values as large as 1000, which greatly complicates
the definition of Leath's  ``active neighborhood'' of a cluster.

Therefore we used a \zk{different} approach, with data structures
typical of algorithms designed for the continuous percolation:
each hypercube is identified by its coordinates, i.e., by $d$ integers,
and the clusters are identified using the union-find algorithm.
With this choice, the computer memory storage, as well as the simulation time of each percolation cluster,
is $\propto (L/k)^d$, which enables one to use large values of $L$ and $k$.
We were able to run the simulations for $L/k\le 150$ ($d=3$), $L/k\le 60$ ($d=4$), $L/k\le 40$ ($d=5$),
$L/k \le 20$ $(d=6)$, and $L/k \le 19$ $(d=7)$, and the maximum values of $L/k$
were limited by the acceptable computation time rather than the storage.

The simulation time in our method is determined by how quickly one can identify all obstacles
connected to the next obstacle being added to the system.
To speed this step up, we divided the system into $(L/k)^d$  bins of linear size $L/k$.
Each obstacle was assigned to exactly one bin and
for each bin we stored a list of obstacles already assigned to it.
In this way, upon adding  a new obstacle, the program had to check $3^d$ neighboring bins
to identify all obstacles connected to the just added one. As $d$ increases, this
step becomes the most time-consuming part of the algorithm.
Fortunately, the negative impact of the $3^d$ factor is to some extent mitigated by the fact that
the critical volume fraction, $\varphi_k^\mathrm{c}$, is much smaller for $k>1$ than for $k=1$,
so that for $k>1$ one needs to generate a relatively small number of obstacles to reach the percolation.
This is related to the fact that for $d\gg 1$ the value of $\varphi_1^\mathrm{c} \approx 1/(2d-1)$
\cite{Grassberger03}, whereas $\varphi_\infty^\mathrm{c} \approx 1/2^d$ \cite{Torquato12b}.
The case $k=1$ is special in that one has to check only $2d$
neighboring sites of a given obstacle, a value much smaller than $3^d$.
Thus, the total simulation time at a high space dimension for a fixed value of $L/k$
can be approximated as $\propto 2d/(2d-1) \approx 1$
for $k=1$ and $\propto (3/2)^d$ for $k > 1$.
In practice, simulations in the space dimension $d=7$ (with $L/k$ fixed)
are between 8 to 13 times faster for $k=1$ than for $k>1$.
This allowed us to run more simulations and obtain more accurate results for $k=1$ than for $k > 1$.

We assumed periodic boundary conditions along all $d$ main directions of the lattice.
The number of independent samples varied from $\approx 3\cdot10^8$ for very small systems (e.g., $d=3$, $k=1$, $L\le10$)
to $\approx 10^4$ for larger $L$ and $d$ (e.g., $d=7$, $k=1$, $L=20$).
Starting from an empty system of volume $L^d$, we added hypercubes of volume $k^d$ at different random locations until we
have detected wrapping clusters in all $d$ directions.
Thus, for each simulation we stored $d$ numbers $n_i$  equal to
the number of hypercubes for which a wrapping percolation was first detected along Cartesian direction $i=1,\ldots,d$.
Having determined all $n_i$ in a given simulation, we can use several definitions of
the onset of percolation in a finite-size system \cite{Newman2001}.
For example, one can assume that the system percolates when there is a wrapping cluster
along some preselected direction $i$, say, $i=1$.
We shall call this definition `case~A'.
Alternatively, a system could be said to be percolating when there is a wrapping cluster
along \emph{any} of the $d$ directions. We shall call this `case~B'.
Another popular definition of a percolation in a finite-size system is the requirement that
the wrapping condition must be satisfied in \emph{all} $d$ directions. This will be denoted as `case~C'.

Next, for each of the three percolation definitions, A, B and C, we determined the probability $P_{L,k}(\varphi)$
that a system of size $L$, obstacle size $k$, and volume fraction $\varphi$
contains a percolating (wrapping) cluster.
This step is based on a probability distribution function
constructed from from all $n_i$ in case A, from all $\min(n_i), i=1,...,d$ in case B, and
from all $\max(n_i), i=1,...,d$ in case C.
Notice that in case A we take advantage of  the symmetry of the system
which ensures that all main directions are equivalent so that
after $N$ simulations we have $dN$ pieces of data from which a single probability distribution function can be constructed.
In doing this we implicitly assume that all $n_i$ are independent of each other, which may improve statistics.

Fixing $d,L$ and $k$, and using (\ref{eq:def-phi}), we can write $P_{L,k}$ as a function of $N$,
$P_{L,k}(\varphi) \equiv P_N$.
In accordance with the finite-size scaling theory,
$P_{L,k}$ is expected to scale with $L$ and the deviation $\varphi - \varphi^\mathrm{c}_k$
from  the critical volume fraction  $\varphi^\mathrm{c}_k$ as \cite{Christensen2005,Rintoul97}
\begin{equation}
  \label{eq:scaling}
  P_N \equiv P_{L,k}(\varphi) = f_k\left(
                         \left[\varphi-\varphi^\mathrm{c}_k\right]L^{1/\nu}
                \right), \quad  L/k \gg 1,
\end{equation}
where $\nu$ is the correlation length exponent,  $f_k$ is a scaling function,
and $\varphi$ is related to $N$ through (\ref{eq:def-phi}).
This formula describes the probability that there is a percolation cluster in a system
containing exactly $N$ obstacles, i.e., is a quantity computed for
a ``microcanonical percolation ensemble''. The corresponding value in the canonical ensemble is \cite{Newman2001}
\begin{equation}
  \label{eq:Q}
  P(p) = \sum_{N=0}^{V} {V \choose N} p^N(1-p)^{V-N}P_N
\end{equation}
where $p$ is the probability that there is an obstacle assigned to a given location;
this  quantity is related to the mean occupied volume fraction through
\begin{equation}
  \label{eq:phi-p}
    \varphi  = 1-(1-p)^{k^d}.
\end{equation}
While $P_n$ is a discrete function, $P(p)$ is defined for all $0\le p \le 1$ and has a reduced statistical noise.
By using (\ref{eq:phi-p}), $P$ can be regarded as a continuous function of $\varphi$ for $0\le\varphi\le 1$.

An effective, $L$-dependent volume fraction $\varphi^\mathrm{c}_k(L)$
was then determined numerically as the solution to
\begin{equation}
  \label{eq:tau}
    P(\varphi) = \tau,
\end{equation}
where $0<\tau<1$ is a fixed parameter, and we chose $\tau=0.5$ in all our calculations.
The critical volume fraction, $\varphi^\mathrm{c}_k$, as well as the critical exponent $\nu$
are then determined using the scaling relation \cite{Oliveira2003}
\begin{equation}
  \label{eq:phi-scaling}
    \varphi^\mathrm{c}_k(L) - \varphi^\mathrm{c}_k =  L^{-1/\nu}(A_0 + A_1 L^{-1} +\ldots+ A_M L^{-M}), \quad L \gg 1,
\end{equation}
where $A_i$ are some $k$- and $d$-dependent parameters and $M$ is the cutoff parameter.

Recently a more general scaling ansatz for the form of the probability $P$
in the vicinity of the critical point was proposed \cite{Wang2013},
\begin{equation}
  \label{eq:U0-general}
    P(\varphi) = U_0 + \sum_{j=1}^{3}q_j(\varphi-\varphi^{\mathrm{c}}_k)^j L^{j/\nu} + \tilde{b}_1 L^{y_i} + \tilde{b}_2 L^{-2},
\end{equation}
where $U_0$ is a universal constant,  $y_i<0$ is the leading correction exponent,
and $\tilde{b}_1,\tilde{b}_2$ are some nonuniversal, model-dependent parameters.
At the critical point this reduces to
\begin{equation}
  \label{eq:U0}
    P(\varphi^{\mathrm{c}}_k) = U_0 + \tilde{b}_1 L^{y_i} + \tilde{b}_2 L^{-2}.
\end{equation}
We used this ansatz to determine the universal, $k$-independent constant $U_0$  \cite{Newman2001,Xu2014,Wang2013}
representing the probability that a wrapping cluster exists at the critical point.

Relation (\ref{eq:phi-scaling}) contains $M+3$ unknowns: $\varphi^\mathrm{c}_k$, $\nu$, and $A_i$.
The critical exponent $\nu$ can be also estimated from an alternative relation
containing $M+2$ unknowns by noticing that  (\ref{eq:scaling}) leads to
\begin{equation}
  \label{eq:derivative-def}
    \left. \frac{\partial f_k}{\partial \varphi}\right|_{\varphi = \varphi^\mathrm{c}_k(L)} =
         L^{1/\nu}f'_k(0) \propto L^{1/\nu}, \quad L\gg1,
\end{equation}
where $f'_k(x) \equiv df_k(x)/dx$. To take into account finite-size corrections, we
used a formula
\begin{equation}
  \label{eq:derivative}
    f'_k(\varphi^\mathrm{c}_k(L)) =
         L^{1/\nu}(B_0 + B_1 L^{-1} +\ldots+ B_M L^{-M}),
\end{equation}
where $B_i$ are some parameters and we used  $M=2$.
Actually, since $f_k(\varphi)$ is a quickly growing function near $\varphi^\mathrm{c}_k$,
we calculated the derivative of its inverse, $f^{-1}_k(\varphi)$ using a five-point stencil,
$(f^{-1}_k)'(x) \approx [-f^{-1}_k(x+2h) + 8f^{-1}_k(x+h) - 8f^{-1}_k(x-h) + f^{-1}_k(x-2h)]/12h$ with $h = 0.001$.

In \cite{Koza14} we conjectured that for sufficiently large $k$
\begin{equation}
  \label{eq:approx-k32}
        1 - \varphi_k^\mathrm{c} -
            \exp\left[
                       \displaystyle
                       \ln (1-\varphi_\infty^\mathrm{c})\frac{ (2k)^d}{(2k-1)^{d-1}(2k + 2d -1)}
               \right]
               \propto k^{-\theta},
\end{equation}
where $\varphi_\infty^\mathrm{c} = \lim_{k\to\infty} \varphi_k^\mathrm{c} $
is the critical volume fraction for the continuous percolation of aligned hypercubes and $\theta = 3/2$.
The left-hand side of (\ref{eq:approx-k32}) was derived using
the excluded volume approximation applied to discrete systems,
whereas its  right-hand-side was obtained numerically and verified for $d=2,3$.
This formula enables one to estimate $\varphi_\infty^\mathrm{c}$
from the critical volumes  $\varphi_k^\mathrm{c}$
obtained for discrete (lattice) models with finite $k$
by investigating the rate of their convergence as $k\to\infty$.
Its characteristic feature is the conjectured independence of $\theta$ on $d$ that
we verify in this report.

The uncertainties of the results were determined as follows.
First, the percolation data were divided into 10 disjoint groups.
Then for each group the value of $\varphi^\mathrm{c}_k(L)$ was calculated in the way described above.
The value of $\varphi^\mathrm{c}_k(L)$ was then assumed to be equal to their average value,
and its uncertainty---to the standard error of the mean.
Next, the value of $\varphi^\mathrm{c}_k$ was obtained from a non-linear fitting
to Eq.~(\ref{eq:phi-scaling}) using the Levenberg-Marquardt algorithm,
with the errors on the parameters estimated from the square roots of the diagonal
elements of the covariance matrix, multiplied by $\max(1, \sqrt{\chi^2/\mathrm{dof}})$,
where $\sqrt{\chi^2/\mathrm{dof}}$ is the reduced chi-square statistic.
The same method was used to estimate the value and uncertainty of $\varphi_\infty^\mathrm{c}$ from Eq.~(\ref{eq:approx-k32}).

Finally, making up the sum in Eq.~(\ref{eq:Q}) is potentially even more tricky than in site percolation,
as in our model $V$ can be as large as \zk{$10^{23}$}.
In solving this technical problem we followed the method reported in \cite{Newman2001} \zk{if $V$ could be stored in a 64-bit integer,
otherwise we approximated the binomial distribution with the normal distribution}.
Another point worth noticing is that $p$ in Eq.~(\ref{eq:Q}) can be as small as \zk{$10^{-14}$}.
In such a case expressions like $(1-p)^{V-N}$ should be computed using appropriate numerical functions, e.g.,
log1p from the C++ standard library, which is designed to produce values of $\log(1+x)$ with $|x| \ll 1$
without a potential loss of significance in the sum $1+x$.


\section{Results\label{Sec:Results}}

\subsection{Site percolation ($k=1$)}
We start our analysis from the particular case $k=1$, in which the model reduces to the standard site percolation.
As site percolation has been analyzed extensively with many dedicated methods,
we were going to use the case $k=1$ only to test the correctness of our computer code,
but as it turned out, we have managed to obtain some new results, too.

The main results are summarized in table~\ref{tab:site-percolation}.
\begin{table}
  \caption{The site percolation threshold $\varphi_1^\mathrm{c}$ and the critical exponent $\nu$
   in dimensions $3\le d \le 7$. Cases A, B, and C refer to three definitions of percolation in a finite-size system,
   whereas I and II refer to two methods of determining $\nu$ (see the text).
   The uncertainty on the last digit(s)
   are given by the figure(s) in the brackets. In the cases marked by an em dash (---) the uncertainties exceeded 100\%.
   The values denoted as ``final'' represent the values obtained for cases A, B, and C combined
   into a single value using the inverse-variance weighting.
   \label{tab:site-percolation}
   }
   \footnotesize
   \begin{tabular*}{\textwidth}{@{}l*{15}{@{\extracolsep{0pt plus 12pt}}l}}
\br
$d$ & case  & \multicolumn{2}{c}{$\varphi_1^\mathrm{c}$}  & \multicolumn{3}{c}{$\nu$}  \\
 \ns
    &       & \crule{2}                                & \crule{3} \\
    &  & best known & present & best known & present (I) & present (II)\\
    \mr
3 & A & 0.311 607 68(15)$^\mathrm{a}$ & 0.311 608 8(57) & 0.876 19(12)$^\mathrm{a}$ & 0.873 6(35) & 0.877 3(12)\\
  & B &                               & 0.311 608 0(42) &                           & 0.855(18)   & 0.874 3(15)\\
  & C &                               & 0.311 601 7(47) &                           & 0.878 7(16) & 0.878 31(80)\\
  \ns\ns\ns
  & & \crule{2} & & \crule{2}\\
  \ns
  &   & \multicolumn{1}{r}{final:}    & 0.311 606 0(48) &                           & \multicolumn{2}{c}{0.877 4(13)}\\
\mr
4 & A & 0.196 886 1(14)$^\mathrm{b}$ & 0.196 890 8(60)  & 0.689(10)$^\mathrm{c}$ & 0.683 0(59) & 0.682 2(41)\\
  & B &                              & 0.196 891 9(55)  &                        & 0.674(18)   & 0.682 7(34)\\
  & C &                              & 0.196 885(10)    &                        & 0.687 9(61) & 0.687 5(23)\\
  \ns\ns\ns
  & & \crule{2} & & \crule{2} \\
  \ns
  & & \multicolumn{1}{r}{final:}     & 0.196 890 4(65)  &                       & \multicolumn{2}{c}{0.685 2(28)}\\
\mr
5 & A & 0.140 796 6(15)$^\mathrm{b}$  & 0.140 796 7(22) & 0.569(5)$^\mathrm{d}$ & 0.574 3(50) & 0.572 5(38)\\
  & B &                               & 0.140 795(10)   &                       & 0.59(11)    & 0.571 6(26)\\
  & C &                               & 0.140 796 5(21) &                       & 0.573 3(29) & 0.572 0(14)\\
   \ns\ns\ns
  & & \crule{2} & & \crule{2} \\
  \ns
  & & \multicolumn{1}{r}{final:}& 0.140 796 6(26)       &                       & \multicolumn{2}{c}{0.572 3(18)}\\
\mr
6 & A & 0.109 017(2)$^\mathrm{b}$  & 0.109 011 3(14) &  1/2 & 0.58(39)  & 0.495(19) \\
  & B &                            & 0.109 017 5(26) &      &   ---     & 0.497(33) \\
  & C &                            & 0.109 009 9(16) &      & 0.513(58) & 0.495(40) \\
   \ns\ns\ns
  & & \crule{2} & & \crule{2}\\
  \ns
  & & \multicolumn{1}{r}{final:}& 0.109 011 7(30) &         & \multicolumn{2}{c}{0.497(25)}\\
\mr
7 & A &  0.088 951 1(9)$^\mathrm{b}$ & 0.088 951 4(56) &   1/2  & 0.36(12)  & 0.44(11) \\
  & B &                              & 0.088 950(15)   &        &  ---      &   ---    \\
  & C &                              & 0.088 945 7(35) &        & 0.40(11)  & 0.41(8)  \\
   \ns\ns\ns
  & & \crule{2} & &  \crule{2} \\
  \ns
  &  & \multicolumn{1}{r}{final:}    & 0.088 951 1(90) &        & \multicolumn{2}{c}{0.41(9)}\\
   \br
  \end{tabular*}
   $^\mathrm{a}$\protect\cite{Xu2014},  \quad
   $^\mathrm{b}$\protect\cite{Grassberger03},  \quad
   $^\mathrm{c}$\protect\cite{Ballesteros97},  \quad
   $^\mathrm{d}$\protect\cite{Christensen2005}.
\end{table} \normalsize
The values of the critical volume fraction, $\varphi_1^\mathrm{c}$, were determined using (\ref{eq:phi-scaling})
with $M=2$ for all three definitions (A, B, and C) of the onset of percolation in finite-size systems,
as defined in section~\ref{sec:numerical-details}. For $d\ge6$ we assumed that $\nu = 1/2$,
whereas for $d < 6$ we treated $\nu$ as an unknown,  fitting parameter.
Percolation thresholds obtained in cases A, B, and C are consistent
with each other and with those reported in other studies \cite{Grassberger03,Xu2014,Stauffer2000}.
We combined them
into a single value using the inverse-variance weighting.
These combined values are listed in table~\ref{tab:site-percolation} as ``final'' values.
Their uncertainty was determined
as the square root of the variance of the weighted mean multiplied by a correction term
$\sqrt{3} \times \max(1, \sqrt{\chi^2/\mathrm{dof}})$, \zk{where $\sqrt{3}$ is an additional, conservative correction term
introduced to compensate for the possibility that the measurements carried out in cases A, B, and C
are not statistically independent, as they are carried out using the same datasets. }

The uncertainties obtained for $d=5,6$ are similar in magnitude to those
obtained with Leath's algorithm~\cite{Grassberger03},
even though our algorithm was not tuned to the numerical features of the site percolation problem.
\zk{It is also worth noticing that the uncertainties of $\varphi_1^\mathrm{c}$ for cases A, B and C are similar
to each other even though case A utilizes a larger number of data. This suggests that the numbers $n_i$
obtained in individual simulations are correlated.
}

The values of the critical exponent $\nu$ were obtained independently using either (\ref{eq:phi-scaling}),
which we call ``method~I'', or (\ref{eq:derivative}) (``method~II'').
Obviously, in contrast to the use of (\ref{eq:phi-scaling}) to estimate
$\varphi_1^\mathrm{c}$, in method I we treated $\nu$ as a fitting parameter for all $d$.
Just as for $\varphi_1^\mathrm{c}$, we publish the values of $\nu$ for individual cases A, B, and C as well as their
combined values obtained with the inverse-variance weighting.
We also present the results for $d=6,7$, where the exact value of $\nu$ is known, as it helps
to verify accuracy of the applied methods.
Again, the results are consistent with the values of $\nu$ reported in previous studies
\cite{Xu2014,Christensen2005,Ballesteros97}.
Method II turned out to be generally more accurate than method I and the accuracy of both methods decreases with $d$.
We attribute the latter phenomenon to a rapid decrease of the maximum system size that can be reached in simulations,
$L_{\mathrm{max}}$, with $d$.
For example, while for $d=3$ we used $L_\mathrm{max} = 200$, for $d=7$ we had to do with $L_\mathrm{max} = 20$,
which certainly has a negative impact on power-law fitting accuracy.
For  $d\ge6$  equation (\ref{eq:phi-scaling}) with $\nu$ treated as a fitting parameter
leads to rather poor fits in which the uncertainty of some fitting parameters may exceed 100\%.
However, the same equation still gives good quality fits after fixing $\nu$ at its theoretical value $1/2$,
which justifies its use for $d\ge 6$ in table~\ref{tab:site-percolation}.

Combining the results from cases A, B, and C and methods I and  II,
we found $\nu = 0.6852(28)$ for $d=4$ and $\nu = 0.5723(18)$ for $d=5$,
which are more accurate than those reported previously,
$\nu = 0.689(10)$ for $d=4$ \cite{Ballesteros97} and $\nu = 0.569(5)$ for $d=5$ \cite{Christensen2005}.
These improved values will be used in data analysis for the case $k>1$.


\subsection{Overlapping hypercubes ($k>1$)}

\begin{figure}
  \includegraphics[width=0.9\columnwidth]{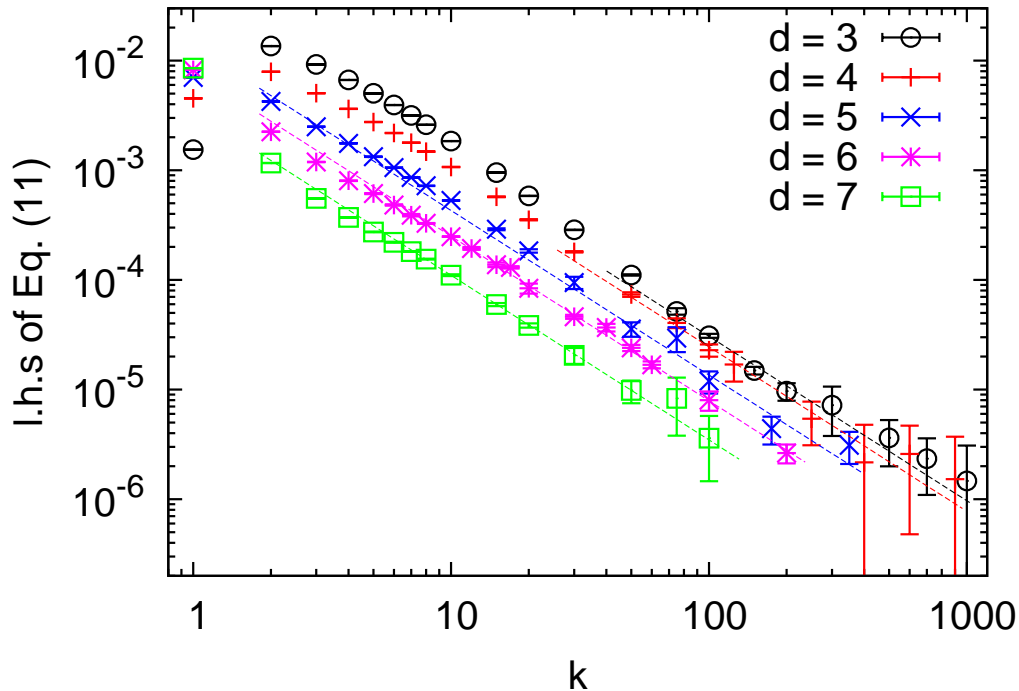}
\caption{\label{fig:exponents-4-7}
 The left hand side of \protect(\ref{eq:approx-k32}) as a function of the obstacle size, $k$.
 Symbols show numerical results for the space dimension $d=3$ (circles), 4 (pluses), 5 (crosses), 6 (stars), and 7 (squares),
 whereas the lines are the best fits to \protect(\ref{eq:approx-k32}) with $\theta=3/2$. }
\end{figure}

Next we verified Eq.~(\ref{eq:approx-k32}) for cases A, B, and C and $3\le d \le 7$.
To this end the values of $\varphi^\mathrm{c}_k$ were determined using (\ref{eq:phi-scaling}) and
$\nu$ fixed at the best value available, i.e., $\nu=0.87619(12)$  for $d=3$ \cite{Xu2014},
our values reported in table~\ref{tab:site-percolation} for $d=4,5$, and $\nu=1/2$ for $d\ge 6$ \cite{Christensen2005}.
The uncertainty of $\nu$ was included into the final uncertainties of the fitting parameters.
Our results, depicted in figure~\ref{fig:exponents-4-7},
confirm our hypothesis that $\theta = 3/2$ irrespective of the space dimension
(similar scaling for $d=2,3$ but percolation defined through spanning clusters was reported in \cite{Koza14}).

This opens the way to use Eq.~(\ref{eq:approx-k32}), with $\theta = 3/2$,
as a means of estimating the continuous percolation threshold of aligned hypercubes, $\varphi^\mathrm{c}_\infty$.
The results, presented in table~\ref{tab:continuous-percolation},
\begin{table}
   \caption{Continuous percolation threshold $\varphi^\mathrm{c}_\infty$ for aligned hypercubes in the space dimension $3\le d \le 7$.
   Cases A, B, and C refer to different definitions of percolation in a finite-size system.
   Also included are the lower ($\varphi^\mathrm{c}_\mathrm{L}$) and upper ($\varphi^\mathrm{c}_\mathrm{U}$) bounds for $\varphi^\mathrm{c}_\infty$,
   calculated from the data reported in \protect\cite{Torquato12b}.
             \label{tab:continuous-percolation}
           }
\begin{indented}
\item[]
\begin{tabular}{llllll}
\br
$d$ & case & \multicolumn{1}{c}{$\varphi^\mathrm{c}_\mathrm{L}$} &  \multicolumn{2}{c}{$\varphi^\mathrm{c}_\infty$} & \multicolumn{1}{c}{$\varphi^\mathrm{c}_\mathrm{U}$} \\
\ns & & & \crule{2} & \\
     &        &                                          & best known        & present                   & \\
    \mr
 3 & A & 0.226 38\ldots & 0.277 27(2)$^\mathrm{a}$ & 0.277 302 0(10)  & 0.293\ldots  \\
   & B &                &                          & 0.277 300 9(10)  &              \\
   & C &                &                          & 0.277 302 61(79) &              \\
 \ns\ns\ns
   &   &                & \crule{2}                                                  \\ \ns
   &   &                & \multicolumn{1}{r}{final:}  & 0.277 301 97(91) \\
 \mr
 4 & A & 0.098 13\ldots & 0.113 2(5)$^\mathrm{b}$  & 0.113 234 40(73) & 0.146\ldots \\
   & B &                &                          & 0.113 233 90(91) & \\
   & C &                &                          & 0.113 237 9(13)  & \\
 \ns\ns\ns
   &   &                & \crule{2} \\ \ns
   &   &                & \multicolumn{1}{r}{final:}  & 0.113 234 8(17)\\
 \mr
 5 & A & 0.043 73\ldots & 0.049 00(7)$^\mathrm{b}$ & 0.048 163 5(15)  & 0.071\ldots  \\
   & B &                &                          & 0.048 165 8(14)  & \\
   & C &                &                          & 0.048 162 1(13)  & \\
 \ns\ns\ns
   &   &                & \crule{2} \\ \ns
   &   &                & \multicolumn{1}{r}{final:}  & 0.048 163 7(19)\\
 \mr
 6 & A & 0.020 03\ldots & 0.020 82(8)$^\mathrm{b}$ & 0.021 347 4(10)  & 0.034\ldots  \\
   & B &                &                          & 0.021 344 6(27)  & \\
   & C &                &                          & 0.021 347 9(10)  & \\
 \ns\ns\ns
   &   &                & \crule{2} \\ \ns
   &   &                & \multicolumn{1}{r}{final:}  &  0.021 347 4(12) \\
 \mr
 7 & A & 0.009 38\ldots & 0.009 99(5)$^\mathrm{b}$ & 0.009 776 9(10)  & 0.017\ldots \\
   & B &                &                          & 0.009 782 0(27)  & \\
   & C &                &                          & 0.009 773 1(10)  & \\
 \ns\ns\ns
   &   &                & \crule{2} \\ \ns
   &   &                & \multicolumn{1}{r}{final:}  & 0.009 775 4(31)\\
    \br
  \end{tabular}
  \newline
  $^\mathrm{a}$\cite{Koza14}.\quad $^\mathrm{b}$\cite{Torquato12b}.
 \end{indented}
\end{table}
turn out far more accurate than those obtained with other methods.
However, they agree with the data reported in \cite{Torquato12b} only for $d \le 4$.
In particular, for $d = 5$ our value of the percolation threshold of aligned hypercubes,
$\varphi^\mathrm{c}_\infty$, is away from the value predicted in \cite{Torquato12b} by $\approx 12\sigma$,
where $\sigma$ is the sum of the uncertainities of $\varphi^\mathrm{c}_\infty$
found in our simulations and that reported in \cite{Torquato12b}.
This indicates that for $d\ge 5$ either our uncertainty estimates or those reported in \cite{Torquato12b} are too small.
This discrepancy is very peculiar, because it exists only for $d\ge 5$,
whereas the results for lower $d$ are in perfect accord.

This finding made us recheck our computations.
The main part of our code is written using C++ templates with the space
dimension $d$ treated as a template parameter.
The raw percolation data is then analyzed using a single toolchain for which $d$ is just a parameter.
This implies that exactly the same software is used for any $d$.
Next, our results for different percolation definitions (A, B, and C) agree with each other well.
Moreover, our results for $k=1$ are in good agreement with all the results available for the site percolation,
and for $k>1$ they satisfy the asymptotic scaling expressed in Eq.~(\ref{eq:approx-k32}).
Also, as shown in table~\ref{tab:continuous-percolation},
all our results for $\varphi^\mathrm{c}_\infty$ lie between the lower and upper bounds,
$\varphi^\mathrm{c}_\mathrm{L}$ and $\varphi^\mathrm{c}_\mathrm{U}$, reported in  \cite{Torquato12b}.
We verified that the reduced chi-square statistic
in practically all fits satisfies $0.5 \le \sqrt{\chi^2/\mathrm{dof}} \le 2$, which indicates that
the data and their uncertainties fit well to the assumed models.
We also implemented the code responsible for the transition from the
microcanonical to canonical ensemble, equation~(\ref{eq:Q}), in such a way that all floating-point
operations could be performed either in the IEEE 754 double (64-bit) or extended precision (80-bit) mode.
The results turned out to be practically indistinguishable, indicating that the code is robust
to numerical errors related to the loss of significance.

\zk{An alternative verification of the results is presented in figure~\ref{fig:verification}.
\begin{figure}
  \includegraphics[width=0.9\columnwidth]{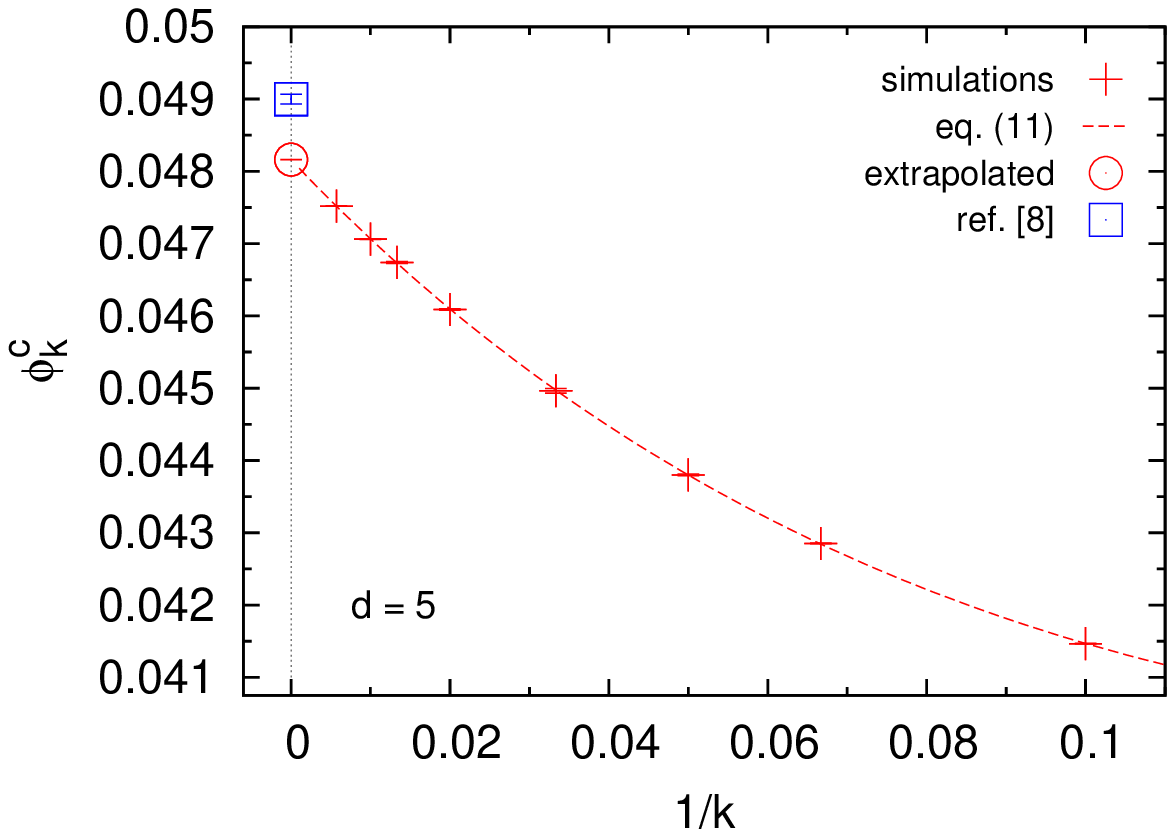}
\caption{\label{fig:verification}
 \zk{Percolation threshold $\varphi_k^\mathrm{c}$ in dimension  $d=5$ for several values of the obstacle linear size $k$.
 Pluses represent our numerical results, the dashed line was calculated
 from a fit to (\protect\ref{eq:approx-k32}) for $k\ge10$,
 the circle depicts the value extrapolated for $k\to\infty$ from (\protect\ref{eq:approx-k32}),
 and the square reproduces  the value of this limit
 as reported in \protect\cite{Torquato12b}.
 }
 }
\end{figure}
It shows that our simulation data for $\varphi_k^\mathrm{c}$ and $d=5$ are in a very good agreement with (\ref{eq:approx-k32}).
The reduced chi-square statistic,  $\sqrt{\chi^2/\mathrm{dof}} \approx 0.9$, indicates a good fit,
even though the uncertainties of individual data points are very small, from $\approx 4\cdot10^{-6}$ ($k=175$)
to $\approx 3\cdot10^{-5}$ ($k=30$)
The value reported in \cite{Torquato12b} for $\varphi_k^\mathrm{c}$ is clearly inconsistent with our data.
The situation for $d=6,7$ is similar (data not shown).
}

It is also worth noticing that our results for $d=3$ are an order of magnitude more accurate than those
 obtained in \cite{Koza14} using exactly the same method, but with the percolation
defined through spanning rather than wrapping clusters. This confirms a known fact that
the estimates of the percolation threshold obtained using
a cluster wrapping condition in a periodic system exhibit significantly smaller
finite-size errors than the estimates made using cluster spanning in open systems \cite{Newman2001}.


\subsection{Corrections to scaling}

One possible cause of the discrepancy between our results for continuous percolation of aligned
hypercubes and those obtained in \cite{Torquato12b} are the corrections
to scaling due to the finite size of the investigated systems.
To get some insight into their role, we used (\ref{eq:U0}) to obtain the values of the universal
constant $U_0$ for $d=3,\ldots,7$ together with $b_1$, $b_2$, $y_i$, which control
the magnitude of the corrections to scaling. First we focused on $y_i$ and found that it
is impossible to find reliable values of this exponent from our data.
Wang et al.\ \cite{Wang2013} also reported difficulties in
determining $y_i$ from simulations, but eventually found $y_i = -1.2(2)$ for  $d=3$.
As the  values of this exponent for $d>3$ are unknown,
and we checked that the value of $U_0$ obtained from  (\ref{eq:U0}) is practically
insensitive to whether one assumes that  $y_i = -1$ or $y_i = -1.2$, we chose the simplest option:
$y_i = -1$ for all $d$, which turns (\ref{eq:U0}) into the usual Taylor expansion in $L/k$,
\begin{equation}
  \label{eq:U0--1}
    P_{L,k}(\varphi^{\mathrm{c}}_k) = U_0 + b_1 (L/k)^{-1} + b_2 (L/k)^{-2},
\end{equation}
where $b_1=\tilde{b}_1k$ and $b_1=\tilde{b}_2 k^2$.

\begin{figure}
  \begin{center}
    \includegraphics[width=0.45\columnwidth]{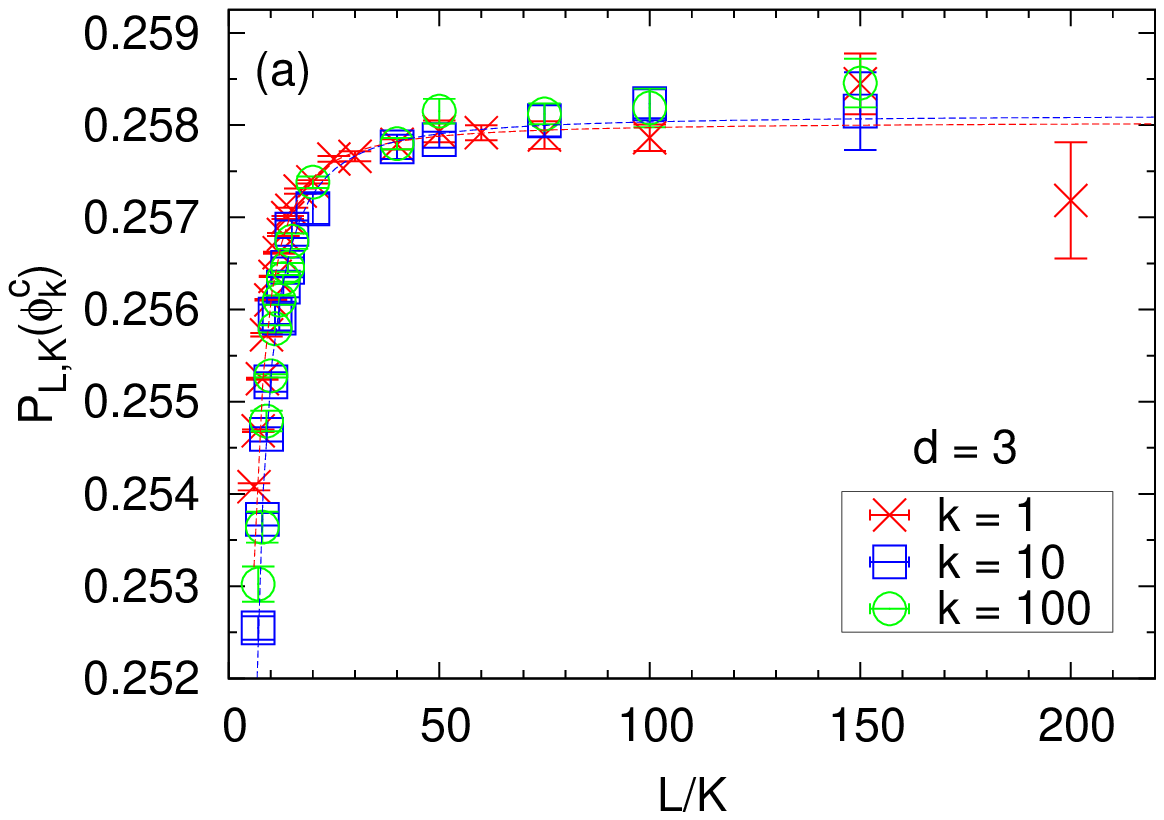}%
    \includegraphics[width=0.45\columnwidth]{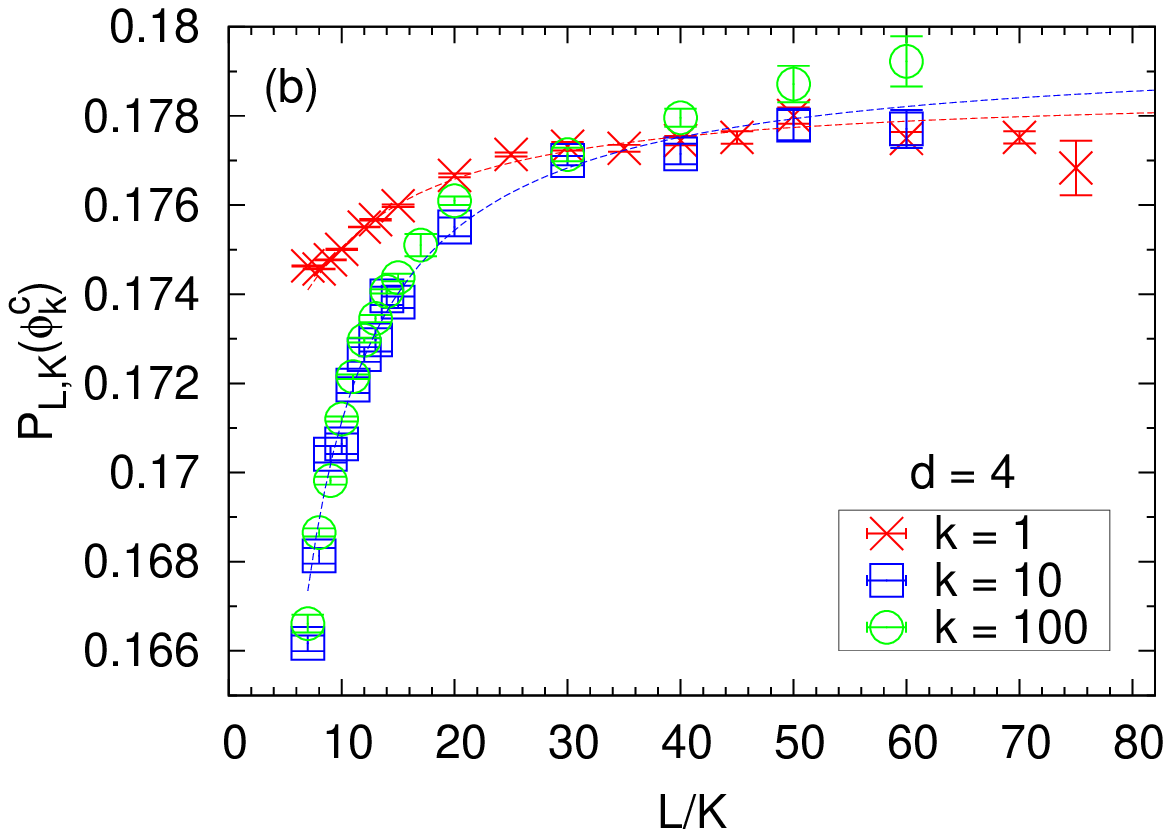}\\
    \includegraphics[width=0.45\columnwidth]{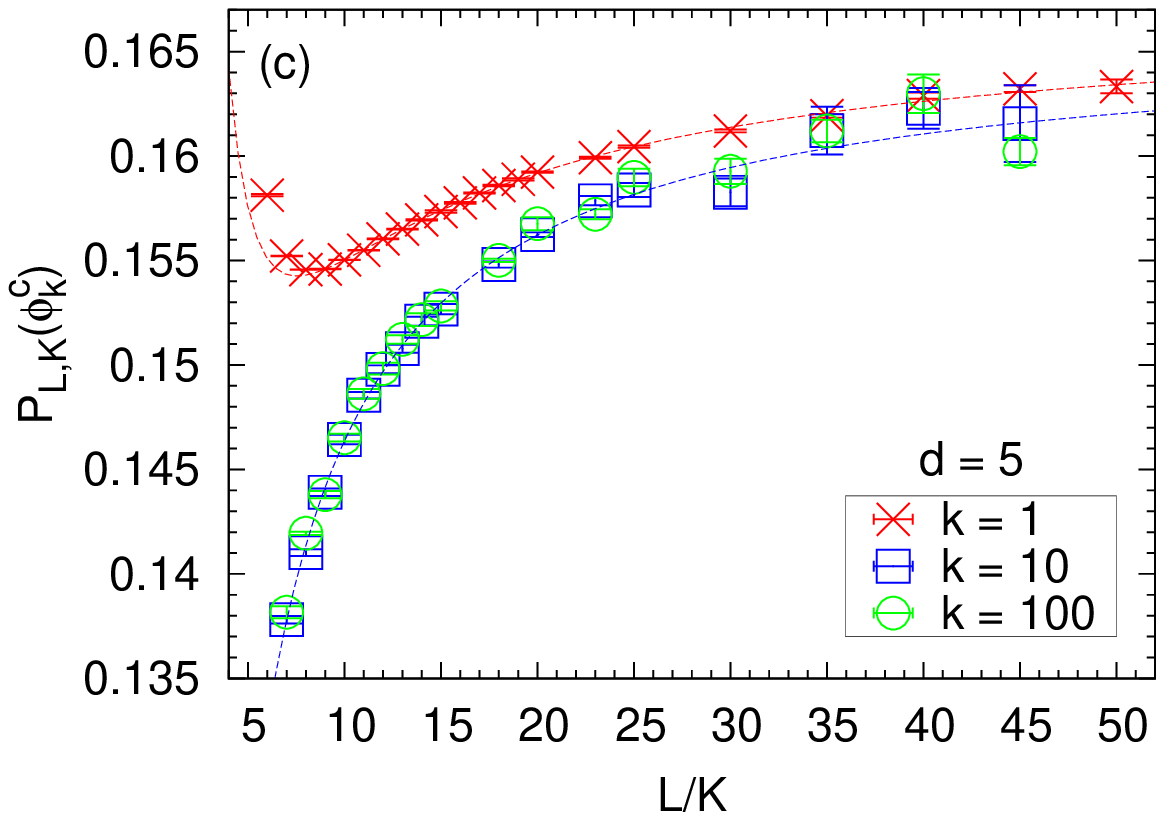}%
    \includegraphics[width=0.45\columnwidth]{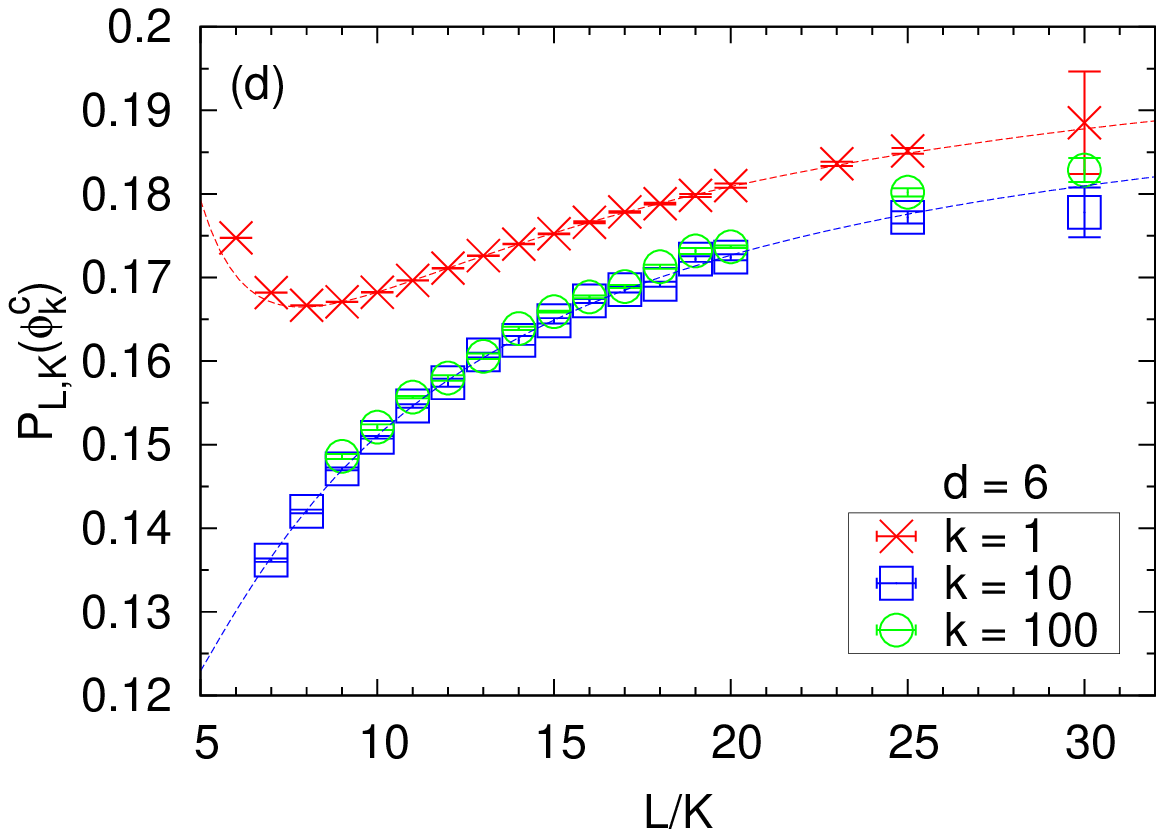}\\
    \includegraphics[width=0.45\columnwidth]{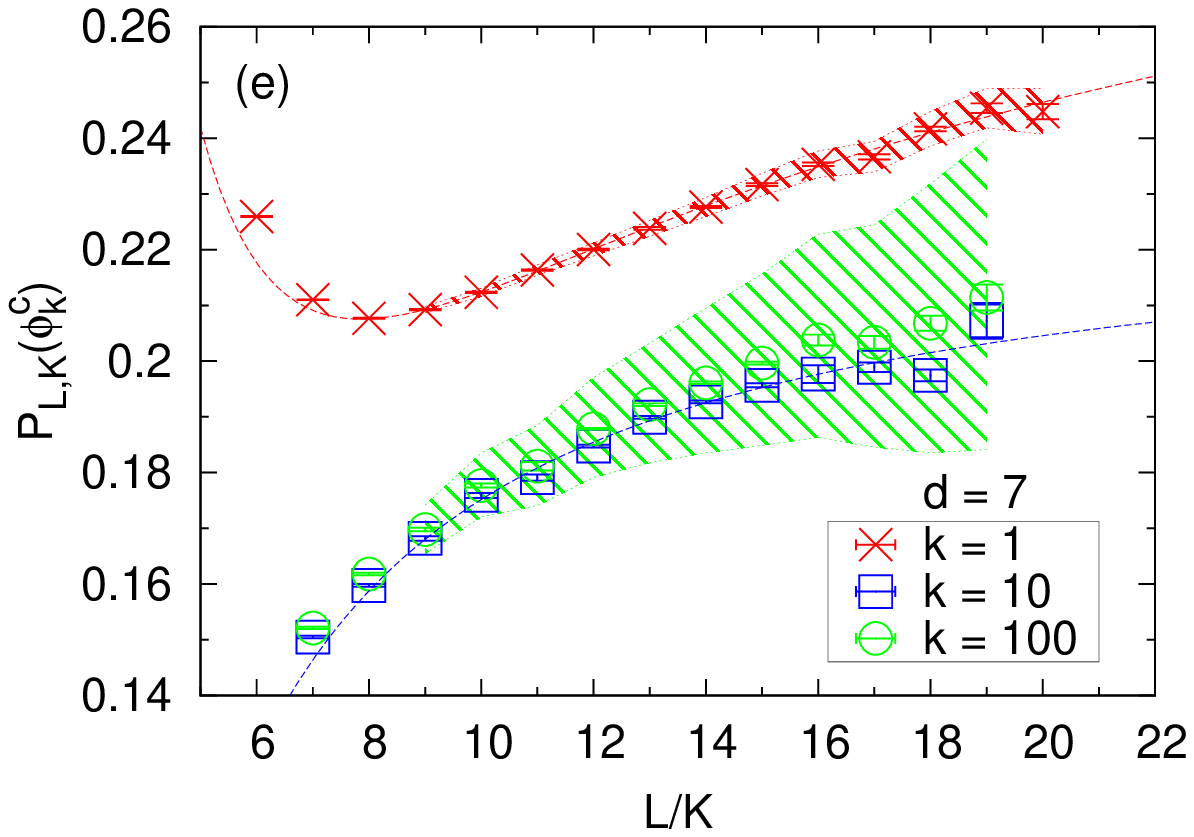}
  \end{center}
  \caption{\label{fig:ypc-individual}
  The probability of a wrapping cluster along a given axis \zk{(case A)} at criticality,
  $P_{L,k}(\varphi^{\mathrm{c}}_k)$, as a function of the system size relative to the obstacle size,
  $L/k$ for selected values of the obstacle size, $k=1$ (red crosses), 10 (blue squares),
  and $100$ (green circles) in dimensions $d=3,\ldots,7$ (panels a,\ldots e, respectively).
  The lines show the fits to (\protect\ref{eq:U0--1}) for $k=1, 10$ (with $L/k\ge9$).
  The lines for $k=100$ would lie very close to those for $k=10$ and are hidden for clarity.
  The error bars do not include the uncertainty of $\varphi^\mathrm{c}_\mathrm{k}$.
  \zk{The regions filled with a pattern in panel (e) show how $P_{L,k}(\varphi^{\mathrm{c}}_k)$ would change if
   the value of $\varphi^{\mathrm{c}}_k$ was allowed to vary by up to its three standard deviations,
   for $k=1$ (red) and $k=100$ (green).
     }
    }
\end{figure}

Figure~\ref{fig:ypc-individual}
shows $P_{L,k}(\varphi^{\mathrm{c}}_k)$ as a function of $L/k$ for $d=3,\ldots,7$ and selected values of $k$ (case A).
As $P_{L,k}(\varphi^{\mathrm{c}}_k)$ is expected to converge to a $d$-dependent limit $U_0$ as $L\to\infty$,
inspection of its convergence rate can serve as an indicator of the magnitude
of the corrections to scaling for the range of the $L$ values used in the simulations.
The plots for $k=10$ are very similar to those obtained for  $k=100$,
which suggests that the behavior of $P_{L,k}(\varphi^{\mathrm{c}}_k)$ for $k=10$ can be used
as a good approximation of $P_{L,k}(\varphi^{\mathrm{c}}_k)$ in the limit of the continuous system, $k\to \infty$.
Rather surprisingly, for $d=3$ this behavior is also similar to that observed in the site percolation ($k=1$).
In higher dimensions the convergence patterns are different: the site percolation is characterized
by a nonmonotonic dependence of $P_{L,k}(\varphi^{\mathrm{c}}_k)$ on $L/k$,
whereas in continuous percolation this dependency is monotonic.
Notice also the different scales used in the plots: the variability of $P_{L,k}(\varphi^{\mathrm{c}}_k)$
increases with $d$ and  at the same time the maximum value of $L/k$ attainable in simulations quickly decreases.
These two factors amplify each other's negative influence on the simulations, which
hinders the usability of the method in higher dimensions.

\zk{Looking at figures~\ref{fig:ypc-individual}~(c)-(e), one might doubt if they represent quantities converging
to the same value irrespective of $k$.}
However, these curves turn out to be very sensitive
to even small changes in $\varphi^{\mathrm{c}}_k$, which are known with a limited accuracy,
a factor not included into the error bars.
\zk{To illustrate the magnitude of this effect, we show in figure~\ref{fig:ypc-individual}~(e)
how $P_{L,k}(\varphi^{\mathrm{c}}_k)$ would change as a function of $L/k$ if $\varphi^{\mathrm{c}}_k$ was
allowed to vary by up to three times its numerical uncertainty
for $d=7$ and $k=1,100$ (case A).
For $k=100$ the impact of the uncertainty of $\varphi^{\mathrm{c}}_k$  on $P_{L,k}(\varphi^{\mathrm{c}}_k)$
turns out larger than the statistical errors, and if we take it into account, the hypothesis that the curves converge
to the same value can no longer be ruled out. Actually, the requirement that this limit is $k$-independent
can be used to argue that our estimation of $\varphi^{\mathrm{c}}_k$ for $d=7$, $k=100$ is larger
than the value obtained from this condition by about twice its numerical uncertainty, which is an acceptable agreement.
While this idea could be used to improve the uncertainty estimates of $\varphi^{\mathrm{c}}_k$ (see \cite{Xu2014}),
we did not use it systematically in the present study. }

\zk{The reason of high sensitivity of $P_{L,k}(\varphi^\mathrm{c}_k)$ to changes in  $\varphi^{\mathrm{c}}_k$
is related to the fact that the slope of $P_{L,k}(\varphi)$ at $\varphi^{\mathrm{c}}_k$
for the largest system sizes attainable in simulations quickly grows with $d$ and, to a lesser extent, with $k$
\begin{figure}
  \begin{center}
    \includegraphics[width=0.9\columnwidth]{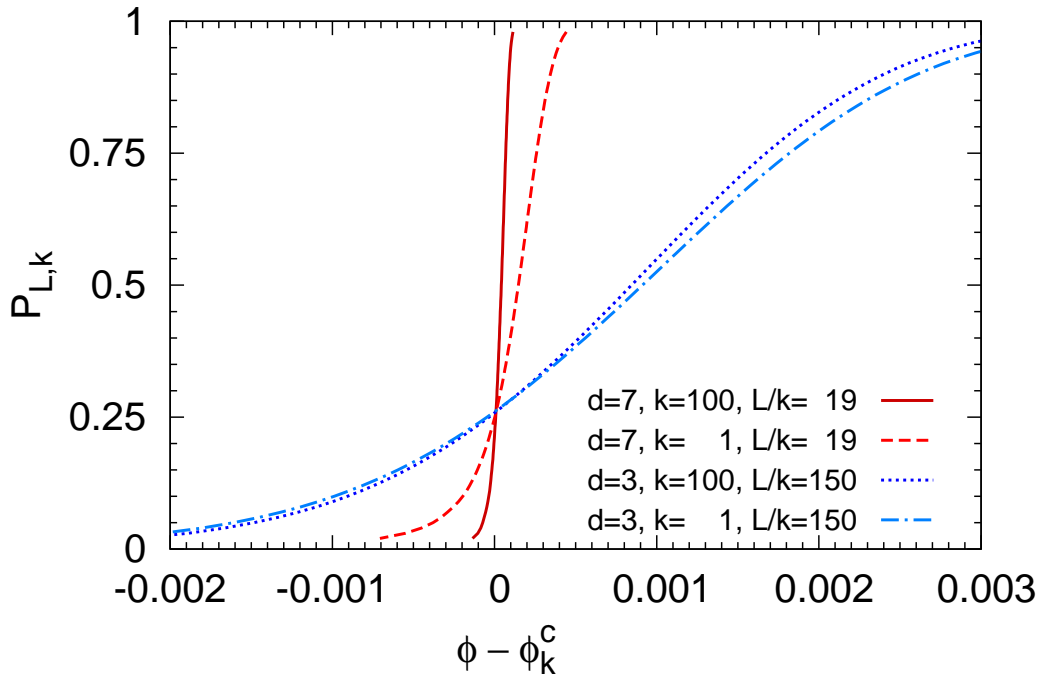}
  \end{center}
  \caption{\label{fig:pdfs}
 \zk{The probability that the system is at percolation, $P_{L,k}$, as a function of the distance to the critical point, $\varphi-\varphi^{\mathrm{c}}_k$,
  for $d=3,7$, $k=1,100$, and the largest values of $L/k$ used in our simulations (case A).}
     }
\end{figure}
(figure~\ref{fig:pdfs}).
For $d=7$ and $k=100$ this slope is as large as $\approx 4600$, so that in this case the uncertainty of $\varphi^{\mathrm{c}}_k$
of the order of $2\times10^{-6}$ translates into the uncertainty of $P_{L,k}(\varphi^{\mathrm{c}}_k)$ of the order of $10^{-2}$.
The data in figure~\ref{fig:pdfs} allows one to make also another observation.
Using (\ref{eq:tau}) with $\tau=0.5$ and the raw data for $d=3$, one can estimate the percolation threshold with the accuracy of $0.001$.
Using extrapolation, this can be improved by a factor of $\approx 1000$ to reach the accuracy reported in table~\ref{tab:continuous-percolation}.
For $d=7$ and $k\ge 10$ the error from the raw data is already very small,  $\approx 4\times10^{-5}$.
Extrapolation can be still used to reduce it further, but since now the data come from systems of
smaller linear size ($L/k\le 19$ rather than $L/k\le 150$ for $d=3$),
the reduction factor is also smaller, of the order of $10$.
Thus, the problems with convergence, which can be seen in panels (c)-(e) of figure~\ref{fig:ypc-individual},
are related to the difficulty in the determination of the universal constant $U_0$, not $\varphi^{\mathrm{c}}_k$.
An independent method of evaluating $U_0$ for $d\ge3$, even with a moderate precision,
would give a powerful method of obtaining the percolation threshold in high dimensional spaces.
}

The values of $U_0$, $b_1$, and $b_2$ obtained from the fits of the data shown
in figure~\ref{fig:ypc-individual} are presented in table~\ref{tab:Rx}.
\begin{table}
   \caption{Parameters of selected fits to equation~(\protect\ref{eq:U0--1}), obtained for $L/k\ge9$ (case A).
   Their uncertainties, shown in the brackets, include the effect of the uncertainty
   of the location of the critical points, $\varphi^{\mathrm{c}}_k$.
             \label{tab:Rx}
           }
\begin{indented}
\lineup
\item[]
\begin{tabular}{cr|lll}
  \br
 $d$  & \multicolumn{1}{c|}{$k$} & \multicolumn{1}{c}{$U_0$} & \multicolumn{1}{c}{$b_1$} & \multicolumn{1}{c}{$b_2$} \\
  \mr
 3 & 1   & 0.2580(2)  & $-0.005(5) $   & $ -0.14(3)$ \\
 3 & 10  & 0.2581(6)  & $-0.006(16)$   & $ -0.22(11)$ \\
 3 & 100 & 0.2583(6)  & $-0.010(14)$   & $ -0.20(9)$ \\
  \mr
 4 & 1   & 0.1786(7)  & $-0.05(2)$     & $ -0.10(7)$  \\
 4 & 10  & 0.1796(19) & $-0.08(5)$     & $ -0.03(31)$ \\
 4 & 100 & 0.1796(16) & $-0.06(4)$     & $ -0.2(2)$ \\
  \mr
 5 & 1   & 0.167(1)   & $-0.19(3)  $   & \m$ 0.7(1)$  \\
 5 & 10  & 0.166(8)   & $-0.19(16) $   & \m$ 0.0(9)$ \\
 5 & 100 & 0.165(9)   & $-0.17(22) $   & $-0.2(12)$ \\
  \mr
 6 & 1   & 0.206(3)   & $-0.62(4)  $   & \m$ 2.4(3)$  \\
 6 & 10  & 0.199(19)  & $-0.58(37) $   & \m$ 1.0(19)$ \\
 6 & 100 & 0.202(18)  & $-0.64(35) $   & \m$1.4(19)$ \\
  \mr
 7 & 1   & 0.313(8)   & $-1.7(1) $     & \m$6.5(8) $ \\
 7 & 10  & 0.229(50)  & $-0.5(10)$     & $-0.9(52)$  \\
 7 & 100 & 0.253(32)  & $-0.9(6) $     & \m$1.2(28)$ \\
  \br
  \end{tabular}
 \end{indented}
\end{table}
Their inspection leads to several conclusions.
First, they agree with the hypothesis that $U_0$ is universal for a given space dimension $d$.
In particular, our value of the universal constant for $d=3$, $U_0 = 0.2580(2)$,
agrees with $U_0 = 0.257 80(6)$ reported in \cite{Wang2013}.
Second, even though the uncertainties of $b_1$ and $b_2$ are typically high, often exceeding 100\% \cite{Wang2013},
one can notice that their magnitude grows with $d$, which means that
the magnitude of the corrections to scaling also grows with $d$.
This is particularly important for $b_1$, which controls the main
contribution to the corrections to scaling for large system sizes $L$.
The absolute value of this parameter  for $k=1$ is very likely to be
at least two orders of  magnitude larger for $d=7$ than for $d=3$.
This translates into much slower convergence of $P_{L,k}(\varphi^{\mathrm{c}}_k)$ for $d=7$ than for $d=3$
(c.f.\ figure~\ref{fig:ypc-individual} and \cite{Stauffer2000}).
Actually, for $d=3$ the value of the linear coefficient, $b_1$, is so close to zero that
the convergence rate of  $P_{L,k}(\varphi^{\mathrm{c}}_k)$ in simulations
is effectively controlled by the quadratic term, $b_2$,
an effect also reported in \cite{Wang2013}.

Once $U_0$ is known with sufficiently low uncertainty,
one can try and use it to reduce the corrections to scaling by assuming $\tau=U_0$ in (\ref{eq:tau}).
This method turned out very successful for  $d=2$ \cite{Newman2001}, but in this case
$U_0$ is known exactly \cite{Pinson1994,Newman2001}.
Availability of the exact value of $U_0$ appears crucial,
because the leading term in (\ref{eq:phi-scaling}) has special properties
only at  $\tau=U_0$ so that small errors in $U_0$ may disturb the fitting.
We checked that, as expected,  setting $\tau=U_0$ in dimensions $d \ge 3$
significantly \zk{increased the convergence rate}; however,
it did not result in more accurate values of the percolation threshold,
probably due to the errors in $U_0$  \zk{and the fact that the uncertainty of
the extrapolated value ($\varphi^\mathrm{c}_\infty$) is closely related to the uncertainty
of the data being extrapolated ($\varphi^\mathrm{c}_k$), which is independent of $\tau$} (data not shown).

Finally, we checked that the value of $U_0$ is universal for other definitions of  percolation in finite-size systems.
If we assumed that a system percolates when a wrapping cluster appears in any direction (case B),
we obtained $U_0^\mathrm{any} = 0.4602(2)$, $0.387(1)$, $0.401(2)$, $0.494(5)$, and $0.659(8)$ for $d=3,\ldots,7$, respectively.
When we waited until a wrapping condition was satisfied along all $d$ directions (case C), we obtained
$U_0^\mathrm{all} = 0.08072(7)$, 0.0291(3), 0.0187(3), 0.0224(9), and 0.043(3) for $d=3,\ldots,7$, respectively.
The values for $d=3$ are consistent with those reported in \cite{Wang2013}, $U_0^\mathrm{any} = 0.459 98(8)$
and $U_0^\mathrm{all} = 0.080 44(8)$.


\section{Conclusions and outlook\label{Sec:Conclusions}}

Treating continuous percolation of aligned objects as a limit of the corresponding discrete model
turned out to be an efficient way of investigating the continuous model.
Using this approach we were able to determine the percolation threshold
for a model of aligned hypercubes in dimensions $3,\ldots,7$ with  accuracy far better
than attained with any other method before. Actually, for $d=4,\ldots,7$
the uncertainty of the continuous percolation threshold is now so small that
it matches or even slightly surpasses that for the site percolation.
We were also able to confirm the universality of the wrapping probability $U_0$ and determine its value for
$d=4,\ldots,7$ for several definitions of the onset of percolation in finite-size systems.

The method proposed here has several advantages.
First, it allows one to reduce the statistical noise of computer simulations
by transforming the results from the microcanonical to canonical ensemble.
Second, it allows to exploit the universality of the convergence rate
of the discrete model to the continuous one, which we found to be controlled by a universal exponent $\theta=3/2$ for all $d$.
Finally, it can be readily applied to several important shapes not studied here, like hyperspheres or hyperneedles.
One drawback of the method is that it does not seem suitable for continuous models
in which the obstacles are free to rotate, e.g. randomly oriented hypercubes.
\zk{We also did not take into account logarithmic corrections to scaling at the upper critical dimension \cite{Stenull2003},
which may render our error estimates at $d=6$ too optimistic.
}

Our results for the continuous percolation threshold in dimensions $d\ge5$
are incompatible with those reported recently in \cite{Torquato12b}.
The reason for this remains unknown, and we guess that they are related to corrections to scaling, which quickly grow with $d$.

Finally, we have managed to  improve the accuracy of the critical exponent $\nu$ measurement in dimensions $d=4,5$.

The source code of the software used in the simulations is available at https://bitbucket.org/ismk\_uwr/percolation.

\section*{Acknowledgments}

The calculations were carried out in the Wroc{\l}aw Centre for Networking and Supercomputing (http://www.wcss.wroc.pl), grant No. 356.
\zk{We are grateful to an anonymous referee for pointing our attention to the Gaussian approximation of the binomial distribution.}


\section*{References}


\end{document}